\documentclass[conference]{IEEEtran}
\IEEEoverridecommandlockouts
\usepackage{cite}
\usepackage{amsmath,amssymb,amsfonts}
\usepackage{algorithmic}
\usepackage{graphicx}
\usepackage{textcomp}
\usepackage{xcolor}
\usepackage{soul}
\usepackage{comment}
\usepackage{hyperref}

\usepackage{eurosym}
\usepackage{amstext} 
\DeclareRobustCommand{\officialeuro}{%
  \ifmmode\expandafter\text\fi
  {\fontencoding{U}\fontfamily{eurosym}\selectfont e}}

\def\BibTeX{{\rm B\kern-.05em{\sc i\kern-.025em b}\kern-.08em
    T\kern-.1667em\lower.7ex\hbox{E}\kern-.125emX}}
\begin{document}

\title{Sharing Mobile and Stationary 
Energy Storage Resources in Transactive Energy Communities\\
\thanks{Support for this research was provided by the Fundação para a Ciência e a Tecnologia (Portuguese Foundation for Science and Technology) through the Carnegie Mellon Portugal Program and by OE - national funds of FCT/MCTES (PIDDAC) under project UID/EEA/00048/2019.}
}

\author{\IEEEauthorblockN{Pedro Moura}
\IEEEauthorblockA{\textit{ISR, Electrical and Computer Eng.} \\
\textit{University of Coimbra}\\
Coimbra, Portugal \\
pmoura@isr.uc.pt}

\and
\IEEEauthorblockN{Uday Sriram}
\IEEEauthorblockA{\textit{Electrical and Computer Eng.} \\
\textit{Carnegie Mellon University}\\
Pittsburgh, USA \\
usriram@andrew.cmu.edu}
\and
\IEEEauthorblockN{Javad Mohammadi}
\IEEEauthorblockA{\textit{Electrical and Computer Eng.} \\
\textit{Carnegie Mellon University}\\
Pittsburgh, USA \\
jmohamma@andrew.cmu.edu}

}

\maketitle

\begin{abstract} 
Most power systems are increasingly based on distributed energy resources, leading to a strong impact on the electrical grid management. In the case of Portugal, the contribution of renewable energy sources to the electricity generation portfolio is already high and the objective is to achieve 100\% by 2050. Most of the new renewable generation capacity will be ensured by distributed photovoltaic generation installed in buildings and therefore the inherent intermittence of photovoltaic output combined with a mismatch with demand profile will challenge the operation and resiliency of the electrical grid. Addressing these issues requires management at the community level and the leveraging of spatio-temporal flexibility of controllable energy resources, such as energy storage resources. This is recognized by the regulators in Portugal and the recent renewable generation self-consumption legislation enables generation-surplus trading in communities. Implementing intra-community trading and utilizing the potentials of renewable generation requires oversight and coordination at the community level in the context of transactive energy systems. This paper focuses on addressing energy sharing through a transactive energy market in community microgrids, using stationary and mobile energy storage as flexibility resources. The proposed framework considers public and commercial buildings with on-site battery storage and numerous electric vehicle charging stations. The formulation is assessed using real data from a community of buildings on a Portuguese University campus. The results showcase that the proposed method achieves an increase in renewable self-consumption at building and community levels, as well as a reduction in electricity costs.
\end{abstract}


\section*{Nomenclature}
\subsection*{Inputs}
\addcontentsline{toc}{section}{Nomenclature}
\begin{IEEEdescription}[\IEEEusemathlabelsep\IEEEsetlabelwidth{$V_1,V_2,V_3$}]
\item[$\Delta h$] Time step ($hour$)
\item[$C_P$]  Baseline parking tariff for EVs parking ($\euro{}/h$)
\item[$C_C(h),$] Tariff for the charging/discharging of EVs at 
\item[$C_D(h)$] \;\ time step $h$ ($\euro{}/h$)
\item[$C_F$] Reward for EV charging flexibility ($\euro{}/h$)
\item[$C_{EG}(h),$] Tariff for power exported/imported to/from 
\item[$C_{IG}(h)$] \;\ the grid at time step $h$ ($\euro{}/kWh$)
\item[$C_{G}(h)$] Tariff for grid use between buildings ($\euro{}/kWh$)
\item[$t^b_{P,n}$] Total parking period of EV $n$ in building $b$ ($h$)
\item[$t^{b+}_{R,n},$] Charging/discharging period requested by EV 
\item[$t^{b-}_{R,n}$] \;\  owner $n$ in building $b$ ($hour$)
\item[$L^{b+}(h),$] Positive/negative net electricity load in building
\item[$L^{b+}(h)$] \;\  $b$ at time step $h$ ($kW$)
\item[$P_{EV,n}^{b+<}$,] Maximum charging/discharging power for EV 
\item[$P_{EV,n}^{b-<}$] \;\ $n$ in building $b$ ($kW$)
\item[$P_{BS}^{b+<}$,] Maximum charging/discharging power for 
\item[$P_{BS}^{b-<}$] \;\ batteries in building $b$ ($kW$)
\item[$\eta^b_{EV,n},$] Efficiency of the charging/discharging of 
\item[$\eta^b_{BS}$] \;\ EVs/batteries in building $b$ ($\%$)
\item[$S_{BS}^{b>},$] Minimum/maximum state of charge of the 
\item[$S_{BS}^{b<}$] \;\ batteries in building $b$ ($\%$)
\item[$E^b_{BS}$] Total capacity of batteries in building $b$ ($kWh$)
\end{IEEEdescription}

\subsection*{Variables}
\addcontentsline{toc}{section}{Nomenclature}
\begin{IEEEdescription}[\IEEEusemathlabelsep\IEEEsetlabelwidth{$V_1,V_2,V_3$}]
\item[$C_E^b(h)$] Electricity cost of building $b$ at time step $h$ ($\euro{}$)
\item[$C^b_{EV}(n)$] Total cost of EV $n$ in building $b$ ($\euro{}$)
\item[$C_{EC}(h),$] Tariff for power exported/imported to/from 
\item[$C_{IC}(h)$] \;\ the community at time step $h$ ($\euro{}/kWh$)
\item[$t^{b+}_{T,n},$] Total charging/discharging period of EV $n$ in 
\item[$t^{b-}_{T,n}$] \;\ building $b$ ($h$)
\item[$t^{b+}_{U,n}(h)$,] Net used charging/discharging period of EV 
\item[$t^{b-}_{U,n}(h)$] \;\ $n$ at time step $h$ in building $b$ ($h$)
\item[$P_{EV,n}^{b+}(h),$] Charging/discharging power of EV $n$ in
\item[$P_{EV,n}^{b-}(h)$] \;\  building $b$ at time step $h$ ($kW$)
\item[$P_{BS}^{b+}(h),$] Charging/discharging power of the batteries 
\item[$P_{BS}^{b-}(h)$] \;\ in building $b$ at time step $h$ ($kW$)
\item[$P^{b+}_c(h)$,] Exporting/importing power flow at time step 
\item[$P^{b-}_c(h)$] \;\ $h$  between building $b$ and community $c$ ($kW$)
\item[$P^\frac{+}{-}(h)$] Ratio between the generation surplus and
\item[$ $] \;\  generation deficit at time step $h$ ($\%$)
\item[$S_{BS}^{b}(h)$] State of charge of batteries in building $b$ ($\%$)
\end{IEEEdescription}

\vspace{0.2cm}
\section{Introduction}
\subsection{Motivation}
The decarbonization and expansion of distributed energy resources are clear drivers of change in the electric power system, which is increasingly based on distributed, intermittent, and non-dispatchable renewable sources. Portugal already has 55\% of the electricity generation ensured by renewables and aims at achieving 100\% renewable electricity generation by 2050 \cite{RNC2050}. This will impact the future of an integrated grid at all scales, but mainly in buildings and communities, since 25\% of the capacity will be ensured by decentralized photovoltaic (PV) generation. However, in most buildings, there is a high mismatch between the local PV generation and demand profiles, leading to the need to export to the grid a significant part of the locally generated energy, even though the same amount is later imported back for local consumption \cite{Vieira2017}. This creates challenges for the electrical grid management and leads to economic losses to the end-user\cite{mohammadi2016distributed}.

In this context, it will be fundamental to have a resilient transactive grid, being the integration and management of new technologies to provide flexibility crucial to achieve such an objective. At building and community levels, distributed energy storage with Battery Storage (BS) systems has emerged as an attractive solution for this new paradigm due to its decreasing costs and increasing efficiency and reliability. Simultaneously, the transport sector with Electric Vehicles (EV) is increasingly an important consumer of electricity and Portugal aims at achieving 70\% electrification of transports by 2050 \cite{RNC2050}. Therefore, EVs parked in buildings can also be used as flexible resources in transactive energy systems, adjusting the charging period with the Building-to-Vehicle (B2V) system, or used as energy storage resources by injecting into the building part of the stored energy with the Vehicle-to-Building (V2B) system \cite{Delgado2018}. Additionally, the new legislation for the self-consumption of renewable generation in Portugal enables the establishment of renewable energy communities, in order to share and trade the renewable generation surplus. Therefore, an aggregated management at the community level will be required to optimize the matching between renewable generation and demand in a transactive energy context.

\subsection{Related Works}
There is a vast body of works proposing methodologies to implement the participation of buildings in transactive energy markets, and the management of flexibility resources.

The implementation of transactive mechanisms for the management of EVs and energy storage is mainly considered in residential buildings. In \cite{Gray2018} a transactive energy control for residential prosumers with BS and EVs is proposed. In \cite{Behboodi2016} the case of EVs participating in a retail double auction electricity regulation market is considered, and in \cite{Liu2019b} a two-stage optimal charging scheme based on transactive control is proposed. However, in residential buildings, the all flexibility resources (including the EVs) belong to the building and there are no economic transactions between entities in order to use such resources. 

Some works have also considered commercial buildings in transactive energy markets. In \cite{Bender2019} the characteristics of commercial buildings and end uses are explored to determine factors supporting the feasibility of participation in transactive energy systems. In \cite{Hao2017} a transactive control market structure for commercial building HVAC systems is presented and in \cite{ Ramdaspalli2016} a passive transactive control strategy was applied to estimate the peak demand reduction potential and energy savings of a building. However, such works only consider demand flexibility, without the use of energy storage resources.

Other works consider the economic relationship between buildings and EV users and. In \cite{Nefedov2018} an office building with PV and EVs is considered with the objective of minimizing energy costs and in \cite{Kuang2017} a building with renewable generation and storage and EVs charging directly with the generated energy is considered in order to ensure the minimizing costs and greenhouse gas emissions. In \cite{Quddus2018} several commercial buildings with EV charging stations are considered with the objective of minimizing the charging costs, as well as the global energy costs in the building. However, such works consider a direct trade of electricity between buildings and EVs with payments associated with the electric power flow which is not allowed for nearly all entities by the actual legislation in most countries. In \cite{Moura2020a}, a first approach based on the parking costs to regulate the economic relationship between building and EV user is proposed, but without considering aggregation at the community level.

In \cite{Liu2019} a transactive real-time EV charging management scheme is proposed for the building energy management system of commercial buildings with PV generation and EV charging. However, such an approach requires complex information from the EV users that is not easily obtained in real scenarios and does not consider the optimization at the community level. In \cite{Moura2020b}, a community is considered, using EVs as flexibility resources, but without considering energy storage and without implementing transactive energy market mechanisms in the aggregation at community level.

\subsection{Contribution}
The main contribution of his work is the design of a transactive energy market for community microgrids constituted by large public and commercial buildings, using BS and EVs as flexibility resources. Therefore, a formulation is proposed to establish a transactive energy market at the community level, using price signals for the energy injected or consumed from the community, in order to give incentives for the aggregated matching between demand and PV generation while ensuring the minimization of electricity costs. Such management is not only ensured with transactions between buildings, but also with flexibility resources in buildings, such as BS and V2B/B2V systems. Therefore, the formulation implements the management of such flexibility resources at the building level. Since the formulation considers the case of large public and commercial buildings with parking lots, where EVs and buildings do not belong to the same entity, a transactive market between buildings and EV users is also established. The economic relationship between EV users and buildings is based on the parking time and added value services for the charging in order to minimize the monitoring requirements and comply with Portuguese legislation that does not allow direct trading of electricity between buildings and EVs.

\subsection{Paper Organization}
The remainder of the paper is structured as follows. Section~II describes the problem formulation and Section~III presents the data and scenarios. The simulation results are presented and discussed in Section~IV. Finally, Section~V summarizes the paper, emphasizing its main conclusions.

\vspace{0.2cm}
\section{Problem Formulation}

\subsection{Objective Function}
The objective of the proposed formulation \eqref{eq:obj} is the minimization of total costs at the community level during the assessed period (i.e., $H$). Therefore, it is considered a community with $B$ buildings, all with PV generation, BS and EVs resources. The total costs in each building take into account not only the electricity costs, but also the income due to the parking and charging of $N$ EVs.

\small
\begin{equation}
\textrm{min}  \sum_{b=1}^{B} \left(\sum_{h=1}^{H}{C^b_E(h)}-\sum_{n=1}^{N}{C^b_{EV}(n)}\right)
\label{eq:obj}
\end{equation}
\normalsize

The electricity cost in each building $b$ \eqref{eq:ce} is associated with the net electricity load and considers the cost/income of energy drawn/injected from the community or from the grid. Therefore, such cost is influenced by the baseline net electricity load in each building, as well as the respective power flows with the BS, EVs and the community.

\vspace{-0.4cm}
\small
\begin{multline}
{C_E^b(h)}= \Delta h \cdot \left[ {P^{b-}_c(h)} \cdot C_{IC} 
+ {P^{b+}_c(h)} \cdot C_{EC}+\right. \\
 \left(L^{b+}(h) - {P^{b-}_c(h)} - {P_{BS}^{b-}(h)} - \sum_{n=1}^{N}{P_{EV,n}^{b-}(h)} \right)  C_{IG}(h) +\\
\left. \left(L^{b-}(h) - {P^{b+}_c(h)} - {P_{BS}^{b+}(h)} - \sum_{n=1}^{N}{P_{EV,n}^{b+}(h)} \right)  C_{EG}(h) \right]
\label{eq:ce}
\end{multline}
\normalsize

The parking and charging service of EVs considers a parking scheme that links the parking and charging periods requested by the EV user, as well as the allowed discharging period with the electricity values, as presented in detail in \cite{Moura2020a}. Therefore, tariffs associated with the parking and charging periods are considered, as well as tariffs to provide rewards by the discharging and flexibility periods. The flexibility tariff ensures a reward for the idle period (i.e., parking period without charging or discharging), since longer idle periods allow higher scheduling flexibility. The discharging period defined by the user is a maximum value and the building can use a lower value. Additionally, the charging period can be higher than the charging period requested by the user in order to compensate for any used discharging period. Therefore, the total parking costs \eqref{eq:ct}, for each EV $n$ in building $b$, depend on the parking period, used periods for charging and discharging in each time step, and the total charging and discharging periods over all time steps.

\vspace{-0.4cm}
\small
\begin{multline}
{C^b_{EV}(n)}=t^b_{P,n} \cdot C^b_P  +  (t^b_{P,n} - t^{b+}_{T,n} - t^{b-}_{T,n} ) \cdot C_F \\
+ \sum_{h=1}^{H} \left( t^{b+}_{U,n}(h)  \cdot C_C(h) \right) +
\sum_{h=1}^{H} \left(t^{b-}_{U,n}(h) \cdot C_D(h)  \right) 
\label{eq:ct}
\end{multline}
\normalsize

Equations \eqref{eq:tcu} and \eqref{eq:tdu} derive the net used charging and discharging periods in each time step and calculate the total periods over all time steps $H$.

\vspace{-0.4cm}
\begin{equation}
\small
t^+_{U,n}(h)= \frac{{P_{EV,n}^+(h)}}{P_{EV,n}^{+<}} \cdot \Delta h, \;\ t^+_{T,n}= \sum_{h=1}^{H} t^+_{U,n}(h)
\label{eq:tcu}
\normalsize
\end{equation}

\vspace{-0.4cm}
\begin{equation}
\small
t^-_{U,n}(h)= \frac{{P_{EV,n}^-(h)}}{P_{EV,n}^{-<}} \cdot \Delta h, \;\ t^-_{T,n}= \sum_{h=1}^{H} t^-_{U,n}(h)
\label{eq:tdu}
\normalsize
\end{equation}

\subsection{Constraints}
The proposed formulation is subject to constraints related to the flexibility resources, as well as with the management of the community.  

\subsubsection{Electric Vehicles}
The periods required by the EV user should be ensured and therefore, the requested charging period, added by the compensation of the discharging period must be achieved until the end of the parking period \eqref{eq:tct}. However, the charging and discharging periods requested by the EV user were defined based on the maximum charging/discharging power. Therefore, such periods include a correction proportional to the ratio between the average and maximum power.

\small
\begin{equation}
t^{b+}_{T,n}=t^{b+}_{R,n}  \frac{P_{EV,n}^{b+<}}{\overline{P}_{EV,n}^{b+}}\color{black}+\frac{t^{b-}_{T,n}}{\eta^b_n}\frac{ \overline{P}_{EV,n}^{b-}}{P_{EV,n}^{b-<}} \label{eq:tct}
\end{equation}
\normalsize

The total discharging period must be lower than the maximum defined by the EV user \eqref{eq:tdm}. Additionally, in each time step, the used discharging period must be lower than the used charging period until such a time step, ensuring that a State of Charge (SoC) lower than the initial value is not achieved. 

\small
\begin{equation}
t^{b-}_{T,n} \leq t^{b-}_{R,n}, \;\ \sum_{h=1}^{x} t^{b-}_{U,n}(h) < \sum_{h=1}^{x} t^{b+}_{U,n}(h) \label{eq:tdm}
\end{equation}

The charging and discharging power is also limited by their maximum values \eqref{eq:ch-limit} and it is not possible to simultaneously charge and discharge the same EV \eqref{eq:chdis}.

\begin{equation}
0\leq {P_{EV,n}^{b+}(h)} \leq {P_{EV,n}^{b++}}, \;\ 0\leq {P_{EV,n}^{b-}(h)} \leq {P_{EV,n}^{b--}}
\label{eq:ch-limit}\\
\end{equation}
\normalsize

\vspace{-0.4cm}
\begin{equation}
{P_{EV,n}^{b+}(h) \cdot P_{EV,n}^{b-}(h)} = 0
\label{eq:chdis}\\
\end{equation}
\normalsize

\subsubsection{Battery Storage}
The charging and discharging power of the batteries is limited by the maximum \eqref{eq:socM} and minimum \eqref{eq:socm} SoC of the batteries \eqref{eq:socbs}, as well as by the maximum charging or discharging power \eqref{eq:pbsm}. Additionally, it is not possible to simultaneously charge and discharge each battery \eqref{eq:pbs}.

\vspace{-0.4cm}
\begin{equation}
\small
{P_{BS}^{b+}(h) \cdot \Delta{h} \cdot \eta^b_{BS,n}} \leq \left(- S_{BS}^{b}(h-1) + S_{BS}^{b<}   \right)  E^b_{BS}
\label{eq:socM}
\normalsize
\end{equation}

\vspace{-0.4cm}
\begin{equation}
\small
{P_{BS}^{b-}(h) \cdot \Delta{h}} \leq \left(S_{BS}^{b}(h-1) - S_{BS}^{b>} \right) E^b_{BS}
\label{eq:socm}
\normalsize
\end{equation}

\vspace{-0.4cm}
\begin{multline}
\small
S_{BS}^{b}(h) =  S_{BS}^{b}(h-1)  
+ \\
\left(\eta^b_{BS}  \cdot  {P_{BS}^{b+}(h)} - {P_{BS}^{b-}(h)}\right) \cdot \frac{ \Delta{h}}{E^b_{BS}} 
\label{eq:socbs}
\normalsize
\end{multline}

\vspace{-0.4cm}
\begin{equation}
\small
{P_{BS}^{b+}(h)} \leq {P_{BS}^{b+<}}, \;\ {P_{BS}^{b-}(h)} \leq {P_{BS}^{b-<}}
\label{eq:pbsm}
\normalsize
\end{equation}

\vspace{-0.4cm}
\begin{equation}
\small
{P_{BS}^{b+}(h) \cdot P_{BS}^{b-}(h)} = 0
\label{eq:pbs}
\normalsize
\end{equation}

\subsubsection{Community}
The import \eqref{eq:pcmax+} or export \eqref{eq:pcmax-} power flow between each building and the community is limited to the net electricity load of such building added by the impact of the flexibility resources. It is also only possible to export to the community if other building needs to import such energy \eqref{eq:pc} and it is not possible to simultaneously import and export to the community \eqref{eq:pb}.

\vspace{-0.4cm}
\small
\begin{equation}
{P^{b-}_c(h)} 
\leq L^{b+}(h) - {P_{BS}^{b-}(h)} - 
\sum_{n=1}^{N}{P_{EV,n}^{b-}(h)} + \sum_{n=1}^{N}{P_{EV,n}^{b+}(h)}
\label{eq:pcmax+}
\end{equation}
\normalsize

\vspace{-0.4cm}
\small
\begin{equation}
{P^{b+}_c(h)} \leq L^{b-}(h) - {P_{BS}^{b+}(h)} - 
\sum_{n=1}^{N}{P_{EV,n}^{b+}(h)}
\label{eq:pcmax-}
\end{equation}
\normalsize

\vspace{-0.2cm}
\small
\begin{equation}
\sum_{b=1}^{B} \left({P^{b+}_c(h)} - {P^{b-}_c(h)} \right) = 0
\label{eq:pc}
\end{equation}
\normalsize

\vspace{-0.2cm}
\small
\begin{equation}
{P^{b+}_c(h) \cdot P^{b-}_c(h)} = 0
\label{eq:pb}
\end{equation}
\normalsize

In order to provide incentives to share the renewable generation surplus in the community, the tariff for exporting energy to the community must between the tariffs of exporting and importing to the grid discounted by the grid use \eqref{eq:ceclim}. The tariff of importing energy from the community should be between the tariff of exporting to the community added by the grid use and the tariff of importing from the grid \eqref{eq:ciclim}. 

\small
\begin{equation}
-C_{EG}(h) \leq -C_{EC}(h) \leq C_{IG}(h)-C_{G}(h)
\label{eq:ceclim}
\end{equation}
\normalsize

\vspace{-0.2cm}
\small
\begin{equation}
-C_{EC}(h)+C_{G}(h) \leq C_{IC}(h) \leq C_{IG}(h)
\label{eq:ciclim}
\end{equation}
\normalsize

The price signals are then adapted for the market conditions by linking the exporting \eqref{eq:cec} and importing \eqref{eq:cic} tariffs with the ratio between the power flow of buildings with generation surplus and the total of the buildings \eqref{eq:pp}, therefore leading to lower/higher prices when the availability of renewable surplus in the community is higher/lower.

\vspace{-0.2cm}
\small
\begin{equation}
C_{EC}(h)=\left(1\hspace{-0.08cm}-\hspace{-0.08cm}P^\frac{+}{-}\hspace{-0.08cm}(h)\right)\hspace{-0.08cm}\left(C_{G}(h)\hspace{-0.08cm}-\hspace{-0.08cm}C_{IG}(h)\right) + P^\frac{+}{-}\hspace{-0.08cm}(h) \cdot C_{EG}(h)
\label{eq:cec}
\end{equation}
\normalsize

\vspace{-0.4cm}
\small
\begin{equation}
C_{IC}(h)=C_{IG}(h)+P^\frac{+}{-}(h) \cdot (C_{G}(h)-C_{EC}(h)-C_{IG}(h))
\label{eq:cic}
\end{equation}
\normalsize

\vspace{-0.2cm}
\small
\begin{equation}
P^\frac{+}{-}(h)=
\frac{\sum_{b=1}^{B} L^{b-}(h)}{\sum_{b=1}^{B} L^{b+}(h)- \sum_{b=1}^{B} L^{b-}(h)}
\label{eq:pp}
\end{equation}
\normalsize

\vspace{0.2cm}
\section{Data and Scenarios}
\subsection{Buildings}
The data used to assess the proposed formulation is from the Department of Electrical and Computer Engineering at the University of Coimbra (Portugal), a building with about 10,000~$m^2$ and yearly electricity consumption of about 500~$MWh/year$. The building is equipped with a PV system with 79~$kWp$, ensuring about 16\% of the actual electricity demand \cite{Fonseca2018}. The actual generation level does not lead to a frequent renewable generation surplus and therefore a future scenario with a larger PV generation system able to ensure 50\% of the demand was considered, being therefore the PV generation data adjusted to the new PV capacity.  

\begin{figure}[htbp]
\centerline{\includegraphics[trim=0 3 10 10,clip, width=0.5\textwidth] {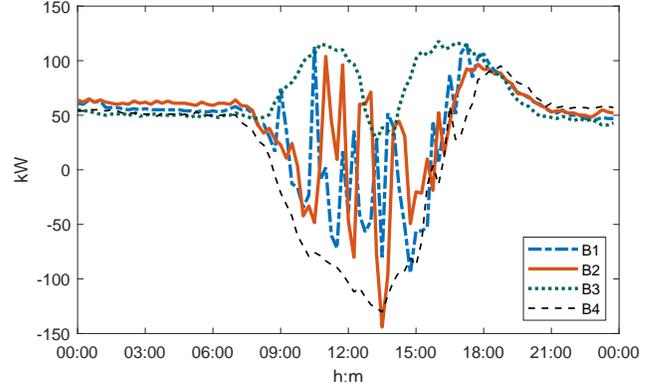}}
\caption{Net electricity load in the four considered buildings}
\label{netload}
\end{figure}

The proposed formulation was simulated for 24 hours periods and in order to ensure scenarios with intermediate consumption and generation, representative days from March were selected. To represent different buildings, with similar characteristics in one community, four different days from different weeks in March were selected to represent the four buildings of the community, as presented in Fig.~\ref{netload}. As can be seen, such data assumes that there are simultaneously buildings with a PV generation surplus and others with a PV generation deficit. In a specific community, the variation of the PV generation presents a high correlation, but the variations of demand may not be correlated, being possible to assume the existence of different net electricity load profiles.

\subsection{Tariffs}
\begin{figure}[htbp]
\centerline{\includegraphics[trim=0 3 10 16,clip, width=0.45\textwidth] {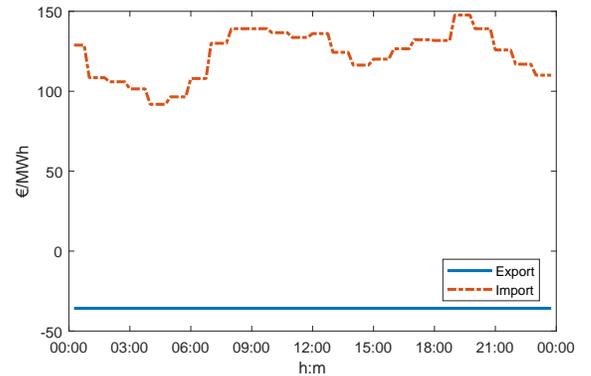}}
\caption{Tariffs for the electricity imported and exported to the grid}
\label{tariffs}
\end{figure}

The actual tariffs in the reference building were used to define the considered tariffs for the electricity imported or exported to the grid. The electricity imported from the grid considers an average cost equal to the average cost in the building (122.8~$\euro{}/MWh$), but with a variation proportional to the average profile of prices in the wholesale market in March. The electricity exported to the grid considers a flat tariff with 90\% of the monthly average of the wholesale market (-35.8~$\euro{}/MWh$), as defined by the Portuguese regulation (Fig.~\ref{tariffs}). In the electricity trading in the community, the considered tariff for the grid use between buildings is a flat tariff of 50~$\euro{}/MWh$.

Regarding the tariffs for the parking and charging of EVs, flat tariffs of 0.5~$\euro{}/h$ and -0.5~$\euro{}/h$, respectively, were considered for the parking and flexibility. The charging and discharging of EVs considered tariffs with a variation proportional to the tariff for the electricity imported from the grid and an average of 2~$\euro{}/h$ and -3~$\euro{}/h$, respectively.

\subsection{Electric Vehicles}
The EVs for the simulations consider a parking period mainly concentrated between 8~$a.m.$ and 8~$p.m.$. As presented in Tab.~\ref{tab:EVs}, requirements of 30 EVs were generated with an average of 8:00~$hours$, 2:00~$hours$ and 0:45~$hours$ for the parking, charging and discharging periods, respectively and a small standard deviation in order to ensure uniform requirements. Such values are aligned with the typical parking requirements in the reference building. From the 30 available EV profiles, six EVs were randomly selected to be parked in each building. The EV chargers used in the buildings considered a maximum charging/discharging power of 10~$kW$ and efficiency of 93\%.

\begin{table}[h]
\centering
\renewcommand{\arraystretch}{1.3}
\caption{Parking requirements}
\label{tab:EVs}
\centering
\begin{tabular}{c c c c c}
\hline
Period & AVG & STD & MIN & MAX\\
\hline
Parking & 8:00 & 1:00 & 6:15 & 11:00\\
Charging & 2:00 & 0:30 & 1:15 & 3:00\\
Discharging & 0:45 & 0:15 & 0:00 & 1:15\\
Start & 9:30 & 0:45 & 8:00 & 10:15\\
\hline
\end{tabular}
\end{table}

\subsection{Battery Storage}
The reference building has a BS system with lithium-ion batteries, ensuring a total storage capacity of about 30~$kWh$ and inverters with a charging/discharging power of 15~$kW$. As in the case of the PV generation, a future scenario was also considered, being used the same adjustment factor. Therefore, a storage capacity of 90~$kWh$ and 45~$kW$ of charging/discharging power was considered.

\vspace{0.2cm}
\section{Simulation Results}
\begin{figure}[htbp]
\centerline{\includegraphics[trim=0 3 10 10,clip, width=0.5\textwidth] {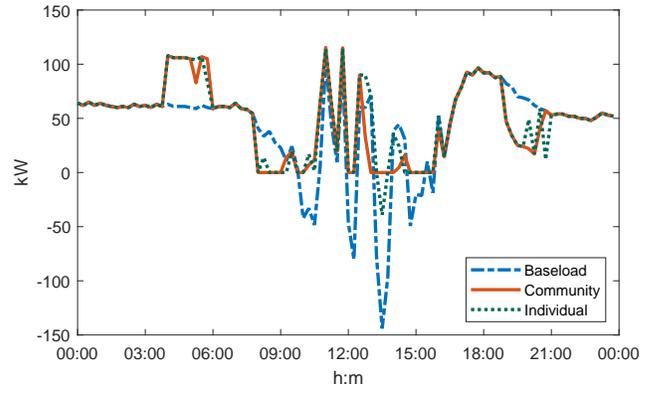}}
\caption{Net electricity load for building 2 with individual and community management}
\label{NLB2}
\end{figure}

\begin{figure}[htbp]
\centerline{\includegraphics[trim=0 0 10 10,clip, width=0.5\textwidth] {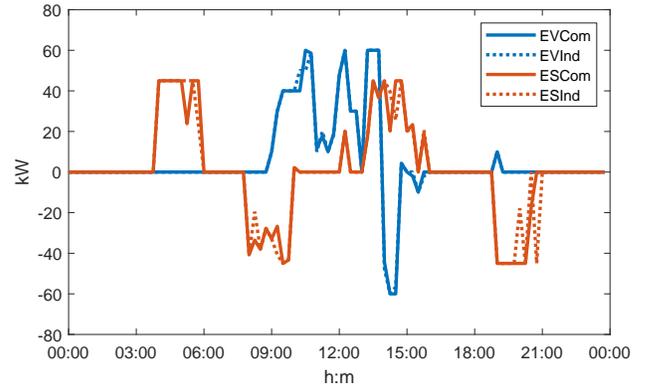}}
\caption{Power flow between EVs, BS and building 2}
\label{EVESB2}
\end{figure}

The formulation was simulated for the presented data and scenarios, considering optimization at building and community levels. For example, Fig.~\ref{NLB2} presents the results for building~2, with the net electricity load for the baseline scenario (without the use of flexibility resources) and scenarios including the impact of EVs and BS with individual and community management of buildings. The use of the flexibility resources in the scenarios with individual and community management is presented in more detail in Fig.~\ref{EVESB2}.

As can be seen, the BS and EVs preferentially charge in periods of negative net electricity load (PV generation surplus) and discharge in periods of positive net electricity load (PV generation deficit). Additionally, the BS also charges during the night taking advantage of the period with lower tariffs for the energy imported from the grid, being such energy partially used to ensure the charging of the first EVs in the morning. The presented net electricity load corresponds to the power flow with the grid, being therefore influenced by the power flow with the community in the case of community management. Therefore, in the case of community management, it was possible to compensate for all periods of negative net electricity load which was not possible with the individual management during a short period. The availability of an additional flexibility resource (the power flow with the community) in the community management justifies the slightly different profiles for the use of BS and EVs.

\begin{figure}[htbp]
\centerline{\includegraphics[trim=0 3 10 10,clip, width=0.5\textwidth] {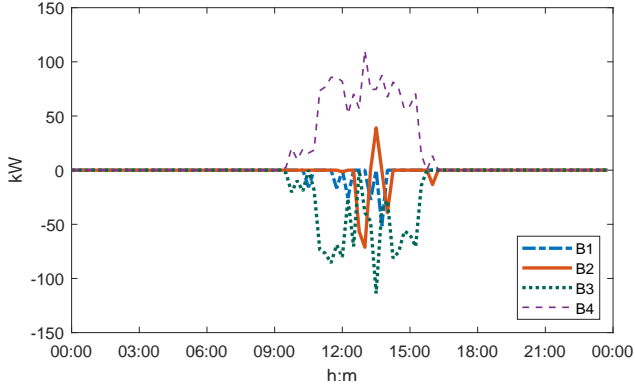}}
\caption{Power flow between each building and the community}
\label{ComPower}
\end{figure}

In the case of community management, there are buildings importing and others exporting to the community, as can be seen in Fig.~\ref{ComPower}. As a result of the market established between such buildings, the tariff for the energy exported to the community was -68.99~$\euro{}/MWh$ and the tariff for the energy imported from the community was 121.48~$\euro{}/MWh$. The costs achieved in the simulated scenarios are presented in Tab.~\ref{tab:Costs}. It should be noted that, when compared with the baseline, the individual and community scenarios require an energy consumption increase of 29.5\% due to the demand associated with the charging of EVs and storage losses. However, due to the use of flexibility ensured by EVs and BS, such higher demand was ensured at a lower cost. Additionally, the management at the community level was able to ensure a 3\% reduction in costs. The objective function takes into account not only the electricity costs, but also the costs paid by EV users, and due to the profit ensured by the designed charging scheme, the total costs relative to the baseline were reduced by 27\%.

\small
\begin{table}[h]
\centering
\renewcommand{\arraystretch}{1.3}
\caption{Costs by building and scenario ($\euro{}$)}
\label{tab:Costs}
\centering
\begin{tabular}{c | r | r r r | r r r}
Buil. & \multicolumn{1}{c|}{Base.} & \multicolumn{3}{c|}{Individual} & \multicolumn{3}{c}{Community} \\
 \# & \multicolumn{1}{c|}{$C_{E}$} & \multicolumn{1}{c}{$C_{E}$} & \multicolumn{1}{c}{$C_{EV}$} & \multicolumn{1}{c|}{Obj.} & \multicolumn{1}{c}{$C_{E}$} & \multicolumn{1}{c}{$C_{EV}$} & \multicolumn{1}{c}{Obj.} \\
\hline
1 & 129.6 & 122.9 & -31.5 & 91.4 & 121.4 & -31.4 & 89.9\\
2 & 140.0 & 141.2 & -39.1 & 102.0 & 139.3 & -39.1 & 100.2\\
3 & 201.9 & 214.5 & -33.0 & 181.5 & 202.3 & -32.8 & 169.5\\
4 & 90.0 & 82.0 & -31.0 & 51.1 & 82.0 & -30.9 & 51.1\\
\hline
Total & 561.5 & 560.6 & -134.6 & 426.0 & 545.0 & -134.3 & 410.7\\
\end{tabular}
\end{table}
\normalsize

\vspace{0.2cm}
\section{Conclusions}
This paper proposes a novel method to establish a realistic transactive energy market for energy sharing in Portugal. The energy trade is envisioned at the community microgrid level and enables mostly by large public and commercial buildings with on-site batteries and EV charging stations. The proposed method also establishes a transactive energy market between buildings and EV users in which parking time and added value services (stemming from charging) are the currency that facilitates economic relationships. The proposed method is aligned with the Portuguese legislation that does not allow direct electricity trading between buildings and EVs while allowing for energy-surplus sharing between buildings in renewable energy communities. 

The formulation was simulated for a building community located at the campus of the University of Coimbra. The results demonstrate that such formulation is able to take advantage of the management of energy storage resources to ensure an increased matching between the demand and local PV generation, as well as lower costs. The proposed method yields a higher impact when the management is implemented at the community level, therefore highlighting the benefits of aggregation ensured by the transactive energy community.

\vspace{0.2cm}
\bibliographystyle{./bibliography/IEEEtran}
\bibliography{./bibliography/IEEEabrv,./bibliography/IEEEexample}

\begin{thebibliography}{10}
\providecommand{\url}[1]{#1}
\csname url@samestyle\endcsname
\providecommand{\newblock}{\relax}
\providecommand{\bibinfo}[2]{#2}
\providecommand{\BIBentrySTDinterwordspacing}{\spaceskip=0pt\relax}
\providecommand{\BIBentryALTinterwordstretchfactor}{4}
\providecommand{\BIBentryALTinterwordspacing}{\spaceskip=\fontdimen2\font plus
\BIBentryALTinterwordstretchfactor\fontdimen3\font minus
  \fontdimen4\font\relax}
\providecommand{\BIBforeignlanguage}[2]{{%
\expandafter\ifx\csname l@#1\endcsname\relax
\typeout{** WARNING: IEEEtran.bst: No hyphenation pattern has been}%
\typeout{** loaded for the language `#1'. Using the pattern for}%
\typeout{** the default language instead.}%
\else
\language=\csname l@#1\endcsname
\fi
#2}}
\providecommand{\BIBdecl}{\relax}
\BIBdecl

\bibitem{RNC2050}
Portugal, ``Roadmap for carbon neutrality 2050,'' Portuguese Republic, Tech.
  Rep., 2019.

\bibitem{Vieira2017}
F.~M. Vieira, P.~S. Moura, and A.~T. de~Almeida, ``Energy storage system for
  self-consumption of photovoltaic energy in residential zero energy
  buildings,'' \emph{Renewable Energy}, vol. 103, pp. 308 -- 320, 2017.

\bibitem{mohammadi2016distributed}
J.~Mohammadi, ``Distributed computational methods for energy management in
  smart grids,'' Ph.D. dissertation, CMU, 2016.

\bibitem{Delgado2018}
J.~Delgado, R.~Faria, P.~Moura, and A.~Almeida, ``Impacts of plug-in electric
  vehicles in the portuguese electrical grid,'' \emph{Transportation Research
  D: Transport and Environment}, vol.~62, pp. 372--385, 2018.

\bibitem{Gray2018}
M.~K. {Gray} and W.~G. {Morsi}, ``A novel transactive energy framework for
  prosumers with battery storage and electric vehicles,'' in \emph{2018 IEEE
  Electrical Power and Energy Conference (EPEC)}, Oct 2018, pp. 1--6.

\bibitem{Behboodi2016}
S.~{Behboodi}, D.~P. {Chassin}, C.~{Crawford}, and N.~{Djilali}, ``Electric
  vehicle participation in transactive power systems using real-time retail
  prices,'' in \emph{2016 49th Hawaii International Conference on System
  Sciences (HICSS)}, Jan 2016, pp. 2400--2407.

\bibitem{Liu2019b}
Z.~{Liu}, Q.~{Wu}, K.~{Ma}, M.~{Shahidehpour}, Y.~{Xue}, and S.~{Huang},
  ``Two-stage optimal scheduling of electric vehicle charging based on
  transactive control,'' \emph{IEEE Transactions on Smart Grid}, vol.~10,
  no.~3, pp. 2948--2958, May 2019.

\bibitem{Bender2019}
S.~R. {Bender}, M.~{Niemeyer}, M.~R. {Weimar}, and T.~D. {Hardy},
  ``Considerations for commercial building participation in a transactive
  energy system,'' in \emph{2019 IEEE Power Energy Society Innovative Smart
  Grid Technologies Conference (ISGT)}, 2019, pp. 1--5.

\bibitem{Hao2017}
H.~{Hao}, C.~D. {Corbin}, K.~{Kalsi}, and R.~G. {Pratt}, ``Transactive control
  of commercial buildings for demand response,'' \emph{IEEE Transactions on
  Power Systems}, vol.~32, no.~1, pp. 774--783, 2017.

\bibitem{Ramdaspalli2016}
S.~{Ramdaspalli}, M.~{Pipattanasomporn}, M.~{Kuzlu}, and S.~{Rahman},
  ``Transactive control for efficient operation of commercial buildings,'' in
  \emph{2016 IEEE PES Innovative Smart Grid Technologies Conference Europe
  (ISGT-Europe)}, 2016, pp. 1--5.

\bibitem{Nefedov2018}
E.~Nefedov, S.~Sierla, and V.~Vyatkin, ``Internet of energy approach for
  sustainable use of electric vehicles as energy storage of prosumer
  buildings,'' \emph{Energies}, vol.~11, no.~8, 2018.

\bibitem{Kuang2017}
Y.~Kuang, Y.~Chen, M.~Hu, and D.~Yang, ``Influence analysis of driver behavior
  and building category on economic performance of electric vehicle to grid and
  building integration,'' \emph{Applied Energy}, vol. 207, pp. 427 -- 437,
  2017.

\bibitem{Quddus2018}
M.~A. Quddus, O.~Shahvari, M.~Marufuzzaman, J.~M. Usher, and R.~Jaradat, ``A
  collaborative energy sharing optimization model among electric vehicle
  charging stations, commercial buildings, and power grid,'' \emph{Applied
  Energy}, vol. 229, pp. 841 -- 857, 2018.

\bibitem{Moura2020a}
P.~{Moura}, G.~{Yu}, S.~{Sarkar}, and J.~{Mohammadi}, ``Linking parking and
  electricity values to unlock potentials of electric vehicles in portuguese
  buildings,'' in \emph{2020 IEEE Power Energy Society General Meeting}, 2020.

\bibitem{Liu2019}
Z.~{Liu}, Q.~{Wu}, M.~{Shahidehpour}, C.~{Li}, S.~{Huang}, and W.~{Wei},
  ``Transactive real-time electric vehicle charging management for commercial
  buildings with pv on-site generation,'' \emph{IEEE Transactions on Smart
  Grid}, vol.~10, no.~5, pp. 4939--4950, Sep. 2019.

\bibitem{Moura2020b}
P.~{Moura}, G.~{Yu}, and J.~{Mohammadi}, ``Management of electric vehicles as
  flexibility resource for optimized integration of renewable energy with large
  buildings,'' in \emph{2020 IEEE PES Innovative Smart Grid Technologies
  Conference Europe (ISGT-Europe)}, 2020.

\bibitem{Fonseca2018}
P.~{Fonseca}, P.~{Moura}, H.~{Jorge}, and A.~T. {de Almeida}, ``Sustainability
  in university campus: options for achieving nearly zero energy goals,''
  \emph{IJSHE}, vol.~19, no.~4, pp. 790--816, Jan 2018.

\end{thebibliography}

\end{document}